\documentclass[aps,prl,reprint,superscriptaddress,amsmath,amssymb]{revtex4-1}

\usepackage{graphicx}
\graphicspath{{./}{./figs/}}
\usepackage{color}
\usepackage[colorlinks, linkcolor=blue, urlcolor=blue, citecolor=blue]{}

\newcommand{\beq}{\begin{equation}}
\newcommand{\eeq}{\end{equation}}
\newcommand{\beqa}{\begin{eqnarray}}
\newcommand{\eeqa}{\end{eqnarray}}

\begin{document}

\title{Spin entangled state transfer in quantum dot arrays: Coherent adiabatic and speed-up protocols}

\author{Yue Ban}
\affiliation{Instituto de Ciencia de Materiales de Madrid, CSIC, C/ Sor Juana In\'{e}s de la Cruz 3, E-28049 Madrid, Spain}
\affiliation{School of Materials Science and Engineering, Shanghai University, 200444 Shanghai, People's Republic of China}

\author{Xi Chen}
\affiliation{Department of Physics, Shanghai University, 200444 Shanghai, People's Republic of China}

\author{Sigmund Kohler}
\author{Gloria Platero}
\affiliation{Instituto de Ciencia de Materiales de Madrid, CSIC, C/ Sor Juana In\'{e}s de la Cruz 3, E-28049 Madrid, Spain}

\date{\today}

\begin{abstract}
Long-distance transfer of quantum states is an indispensable part of
large-scale quantum information processing. We propose a
novel scheme for the transfer of two-electron entangled states, from one edge of a quantum dot array to the
other by coherent adiabatic passage. This protocol is mediated by pulsed tunneling barriers. In a second step, we seek for a speed up by shortcut to adiabaticity techniques.  This significantly reduces the operation time
and, thus, minimizes the impact of decoherence.  For typical parameters of
state-of-the-art solid state devices, the accelerated protocol has an
operation time in the nanosecond range and terminates before a major
coherence loss sets in.  The scheme represents a promising candidate for
entanglement transfer in solid state quantum information processing.
\end{abstract}

\maketitle

Semiconductor quantum dots (QDs) with long spin coherence times, are
promising platforms for quantum information processing.
Scaling up these devices requires techniques for long-range qubit control,
for which several methods and techniques have been proposed and
implemented.  For example, spin circuit quantum electrodynamics
architectures with spin qubits coherently coupled to microwave frequency
photons allow photon-mediated long-distance spin entanglement
\cite{cavity-spin1,cavity-spin2,cavity-spin3,cavity-spin4,david}. Surface
acoustic waves can capture and steer electrons \cite{SAW1} between
distant QDs and, thus, transfer quantum information \cite{SAW2}.

Already triple quantum dots (TQDs) in series allow to investigate
transfer between sites that are not directly coupled. Experimental
evidence of direct electron spin transfer between the edge dots
\cite{quantum-coherence1,quantum-coherence2} and of photo-assisted long
range electron transfer
\cite{photo-cotunneling1,photo-cotunneling2,photo-cotunneling3,photo-cotunneling4}
demonstrates new transfer mechanisms mediated by virtual transitions.
Recent experiments with larger QD arrays \cite{van4,tar4,tar5,
PuddyAPL2015,12dots-Petta} put the challenge of transferring quantum
states over larger distances.  Since such transfer obviously is more
involved, proposals for fast and reliable protocols are particularly
welcome.

Adiabatic transfer has been widely invoked for quantum information
processing. A single electron in a TQD can be directly transferred between
outer dots by adiabatic passage via a dark state (DS) without ever populating
other instantaneous eigenstates of the system \cite{CTAP-QD,sigmund}.
Such techniques, known as coherent transfer by adiabatic passage (CTAP),
have been extended to more complicated architectures in solid state devices \cite{CTAP-QD-chain1,CTAP-QD-chain2,CTAP-spin1,CTAP-exchange-spin-qubit} as well as in cold atoms
\cite{CTAP-atom-transport,CTAP-atom-entanglement}.  Other proposals
consider the level detuning as a control parameter to transfer a qubit
state \cite{ST-detuning-TQD}.  Recent experiments enable coherent shuttling
of electron spin entangled states in QD arrays
\cite{Vandersypen-2017,Tarucha-2018}, where one party of an entangled spin
pair is adiabatically  transferred to a long-distant site by detuning
levels. Also shuttling in parallel two and three electrons at a time has
been implemented in silicon QD arrays \cite{petta}.

\begin{figure}[b]
\centerline{\includegraphics[width=\columnwidth]{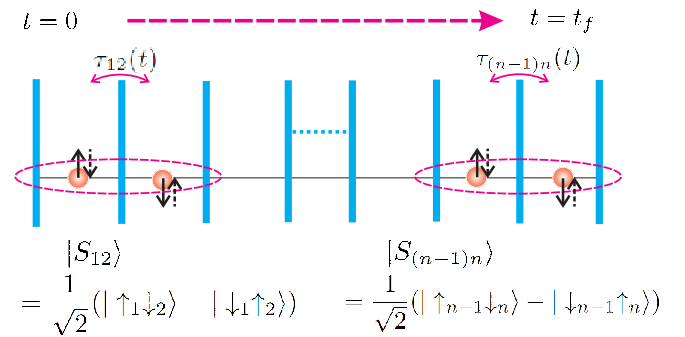}}
\caption{\label{3dots} Direct coherent transfer of a
singlet between outer dots in an array of $n$ QDs.
With the application of pulses $\tau_{12}(t)$, \ldots,
$\tau_{(n-1)n}(t)$, high-fidelity transfer of the singlet state
$|S_{12}\rangle$ to $|S_{(n-1)n}\rangle$ is achieved.
The subscripts refer to the dots occupied by one electron.
}
\end{figure}

In this Letter, 
we propose a new protocol which allows for transferring two-particle entangled states by applying pulses in an adiabatic
manner. 
We focus on the transfer of a singlet
and a triplet
from the first two dots of an array to the last two dots, see
Fig.~\ref{3dots}.  Interestingly, in contrast to the case of a single
particle \cite{CTAP-QD,QDarrays-STA}, we will find that CTAP of an
entangled pair can be achieved also in arrays with an even number of dots.
Furthermore, we improve our protocol by means of shortcuts to adiabaticity
(STA) \cite{STA}, which results in a significant speed-up that reduces the
impact of decoherence.  The feasibility is underlined by a numerical study
that considers charge noise as well as fluctuations of the pulse
intensities.

\begin{figure}[tb]
\begin{center}
\scalebox{0.44}[0.44]{\includegraphics{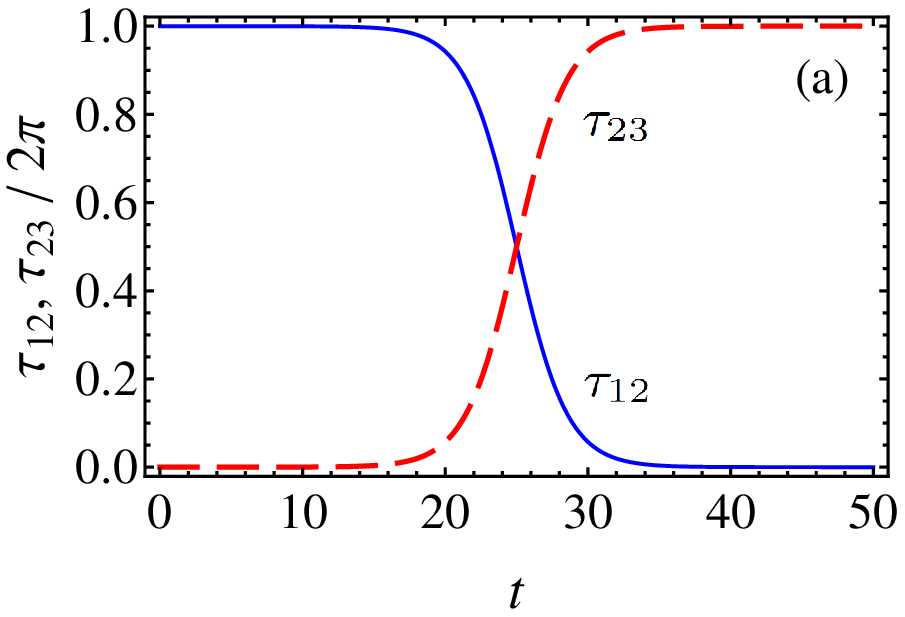}}
\scalebox{0.44}[0.44]{\includegraphics{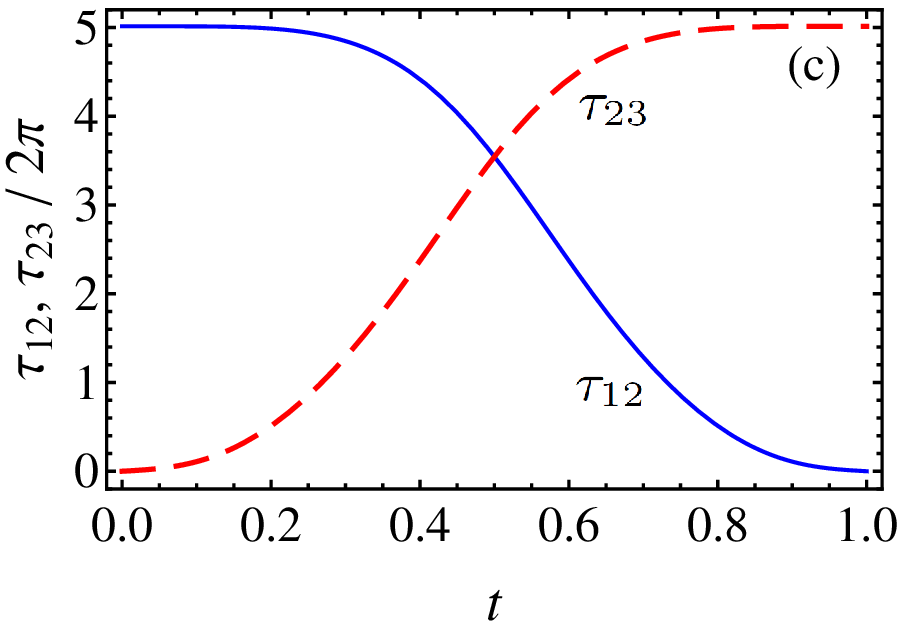}}
\\
\scalebox{0.44}[0.44]{\includegraphics{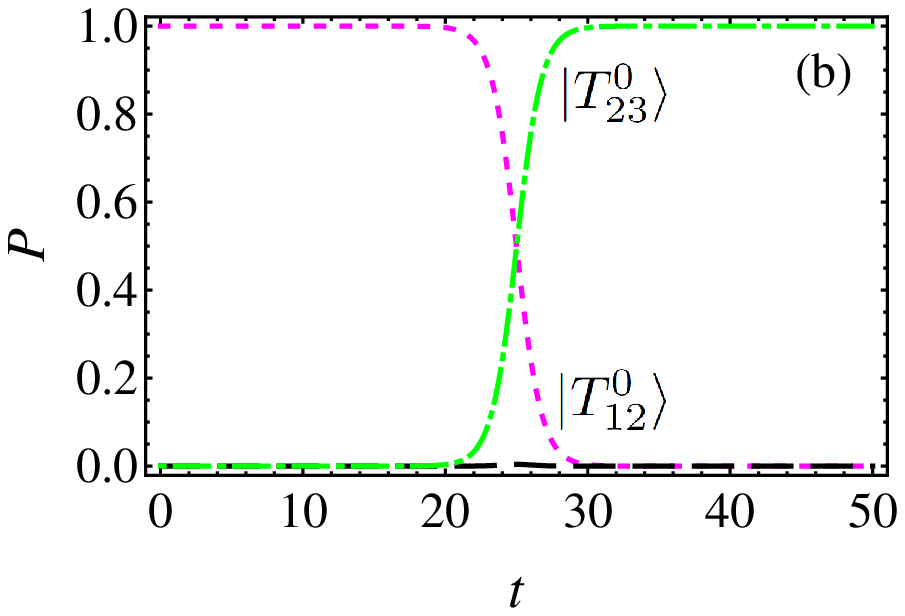}}
\scalebox{0.44}[0.44]{\includegraphics{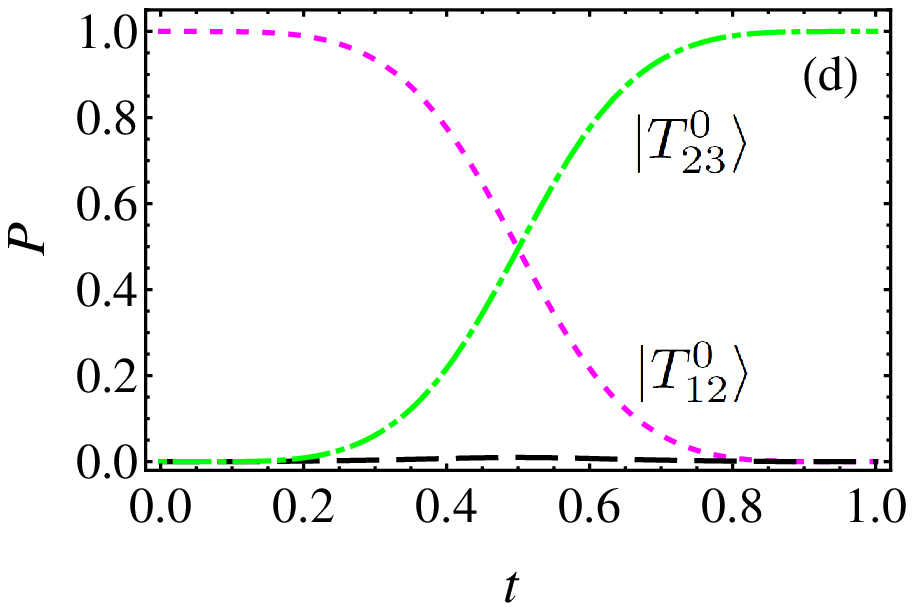}}
\caption{\label{TS-3dots}%
Transfer of a spin triplet in a TQD with a CTAP protocol with duration
$t_f=50$ (left column) and with a STA protocol with $t_f=1$ (right column).
Time in units of $2\pi/\tau^\text{max}_{ij}$.
(a) CTAP pulses $\tau_{12}$ and $\tau_{23}$ controlling
the tunneling between neighbor dots [see Eq.~\eqref{3dots-pulses}].
(b) Time evolution of the populations for the transition
$|T^0_{12}\rangle \to |T^0_{23}\rangle$, where $|T^0_{13}\rangle$ remains
unpopulated.
(c) Pulses of the STA protocol [Eq.~\eqref{pulses-STA}]
and (d) the resulting populations.
}
\end{center}
\end{figure}

\textit{Hamiltonian and Dark State.}---%
We consider two electrons confined in an  array
of $n$ QDs described by the Hubbard model, in the presence of an external
magnetic field $B$ ($\hbar=1$),
$ H = H_\varepsilon + H_\tau + H_U+ H_B$, where
$H_\varepsilon = \sum_{j,\sigma} \varepsilon_j n_{j, \sigma}$ with
$n_{j, \sigma} = c^\dag_{j\sigma} c_{j \sigma}$ the occupation of dot $j$
with spin $\sigma$ and onsite energy $\varepsilon_j$ and $H_\tau = \sum_{<i,j>,\sigma} \tau_{ij} c_{i\sigma}^\dag c_{j\sigma}+ h.c.,$
with tunnel barriers $\tau_{ij}$. Coulomb interaction
$H_U = U_0 \sum_j n_{j,\uparrow} n_{j,\downarrow}$ and $B$ yields $H_B = \sum_{j,\sigma}\Delta_j S^z_{j,\sigma}$
with the Zeeman splittings $\Delta_j$.  For a sufficiently large $B$, the states with parallel spins are energetically far off
and can be neglected.  Then only the triplets with antiparallel spins and the singlets
are relevant.

\begin{figure}[tb]
\begin{center}
\scalebox{0.45}[0.45]{\includegraphics{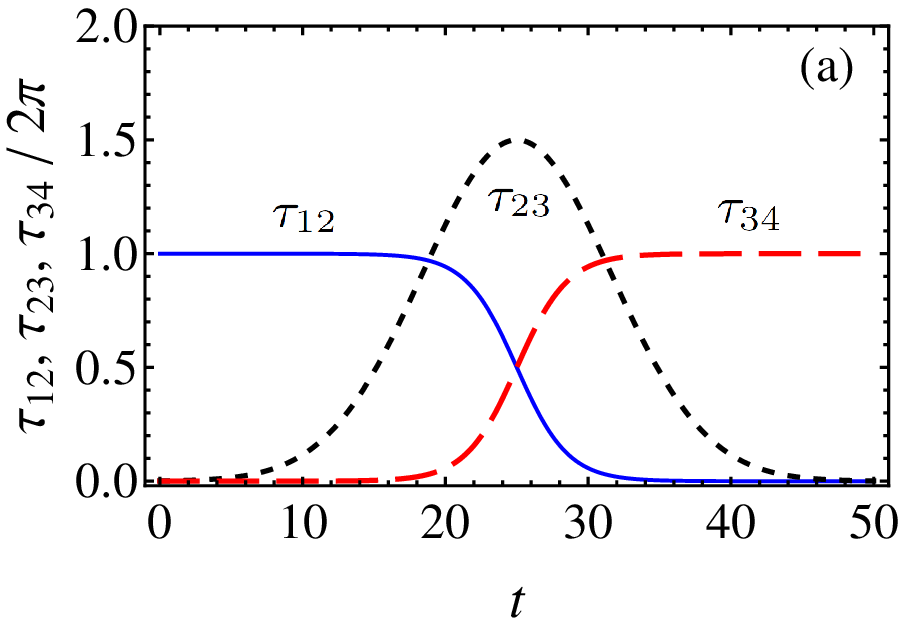}}
\scalebox{0.45}[0.45]{\includegraphics{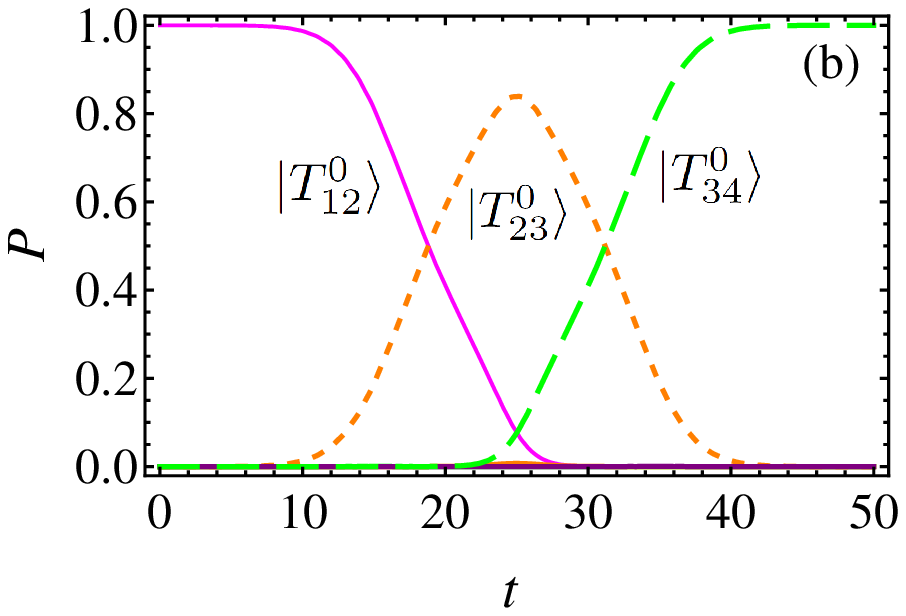}}
\caption{\label{R-CTAP-4dots-1} CTAP transfer of the triplet state in a
QQD with $t_f = 50$. (a) With the application of the pulses $\tau_{12}$, $\tau_{23}$ and $\tau_{34}$,
see Eq.~\eqref{4dots-pulses}, (b) $|T^0_{12}\rangle$ is transferred to $|T^0_{34}\rangle$. During the
transfer there is a finite occupation of the intermediate dots $|T^0_{23}\rangle$.}
\end{center}
\end{figure}

Owing to the Pauli principle, the triplets within the same QD can be formed
only by electrons in different orbital levels and they are largely detuned.  Thus, the triplet basis consists of
$N = \frac{1}{2}(n-1)n$ states $|T^0_{ij}\rangle =
(|{\uparrow}_i{\downarrow}_j\rangle +|{\downarrow}_i{\uparrow}_j\rangle) /
\sqrt{2}$, where the subscripts $i, j \leq n$ refer to the dots occupied.
An important ingredient for CTAP is the emergence of a DS, i.e.,
an eigenstate of the Hamiltonian with zero energy at all times.
In the triplet subspace, it reads \cite{supple}
\beqa
\label{dark-state}
|\phi^T_{DS,1}\rangle
=\frac{1}{|\nu_1|}\sum_{j=1}^{n-1}(-1)^{j+1} \tau_{j(j+1)} |T^0_{{j(j+1)}}\rangle,
\eeqa
where $\nu_1^2 = \sum_{j=1}^{n-1}\tau^2_{j(j+1)}$. Its interesting feature
is that the occupation of a triplet $|T^0_{j(j+1)}\rangle$ is fully
determined by the values of $\tau^2_{j(j+1)}/\nu_1^2$. Therefore, they can be controlled by adiabatically switching $\tau_{j(j+1)}$ such that $|\phi^T_{DS,1}\rangle$ turns from the initial
state $|T^0_{12}\rangle$ to the desired final state $|T^0_{(n-1)n}\rangle$
\cite{CTAP-QD, MenchonEnrich2016a}.

For the singlets, we have $N$ single-occupancy states, but in an
anti-symmetric spin state, namely $|S_{ij}\rangle =
(|{\uparrow}_i{\downarrow}_j\rangle
-|{\downarrow}_i{\uparrow}_j\rangle)/\sqrt{2}$. In addition, there are
$n$ double occupied states (DOST), $|S_{ii}\rangle =
|{\uparrow}_i{\downarrow}_i\rangle$ with energy of the order
$U_0$.  It can be expected that for $U_0 \gg |\tau^\text{max}_{ij}|^2$, with
$\tau^\text{max}_{ij}$ the maximal value of the pulses, both singlet subspaces are
energetically separated such that no transition between them occurs.  Then
one finds again the DS in Eq.~\eqref{dark-state}, but with $T^0$
replaced by $S$ and accordingly denoted by $|\phi^S_{DS,1}\rangle$.


\textit{CTAP Transfer of entangled states in a TQD.}---
To work out the principles of the transfer, we start with a TQD in which
the initial and the final states are $|T^0_{12}\rangle$ and
$|T^0_{23}\rangle$, respectively, while the DS reads
$|\phi^T_{DS,1} \rangle = \cos\theta |T^0_{12}\rangle - \sin\theta
|T^0_{23}\rangle$, where $\tan\theta =\tau_{23}/\tau_{12}$.
Setting $\varepsilon_j=0$ for all $j$, we employ pulses of the form
\begin{equation}
\label{3dots-pulses}
\tau_{12}(t) = -\tau_0 \left[\tanh\left(\frac{t-b}{c}\right) - 1 \right]
= 2\tau_0 - \tau_{23}(t)
\end{equation}
with the pulse parameters $b= t_f /2$ and $c = t_f/14$ [see
Fig.~\ref{TS-3dots}(a)]
.  These pulses are chosen such that the boundary conditions $\theta(0) = 0$ and $\theta(t_f) =
\pi/2$ are fulfilled and, thus, the initial state $|T^0_{12}\rangle$ turns into
the final state $|T^0_{23}\rangle$. We choose the operation time $t_f = 50$ (in units of
$2\pi/\tau_{ij}^\text{max}$), with the maximal intensity of the pulse
$\tau^\text{max}_{ij}=2 \tau_0 = 2 \pi$, such that the adiabaticity condition
$\tau^\text{max}_{ij} t_f = 100 \pi \gg 1$ \cite{CTAP-QD} is fulfilled.  For
QDs, a possible experimental value is $\tau^\text{max}_{ij}=5\mu$eV, such that $t_f=260$ns.
Figure~\ref{TS-3dots}(b) illustrates how the state $|T^0_{12}\rangle$
is adiabatically transferred to $|T^0_{23}\rangle$ along the DS, while
$|T^0_{13}\rangle$ remains unpopulated.  The computed fidelity of
this process for $t_f = 50$ is $F = |\langle T^0_{23}| \Psi(t_f) \rangle|^2
\gtrsim 0.999$.

For the singlet case, we have to include DOST. With the same coupling pulses $\tau_{12}$ and $\tau_{23}$ [Eq. (\ref{3dots-pulses})], the singlet state can be transferred with high fidelity from $|S_{12}\rangle$ to $|S_{23}\rangle$, for $U_0\gg |\tau_{ij}^{\textrm{max}}|^2$. In this case, adiabatic elimination of DOST in the Hamiltonian $H_S$ can be applied
\cite{supple}. For small $U_0$, we find that the fidelity $F = |\langle S_{23}| \Psi(t_f)
\rangle|^2$ of the protocol oscillates heavily.  This is characteristic for
$U_0$ such that DOST are energetically close to the
single occupied ones.  Then the DS $|\phi^S_{DS,1}\rangle$ is no longer
an instantaneous eigenstate.  For the more realistic $U_0\gtrsim 1400\pi \gg
\tau_0$ ($1400\pi$ corresponds to 3.5\,meV), however, the energetic separation of the two
singlet subspaces is sufficiently large such that the transfer fidelity for both singlets and triplets is similar.

\begin{figure}[tb]
\begin{center}
\scalebox{0.5}[0.5]{\includegraphics{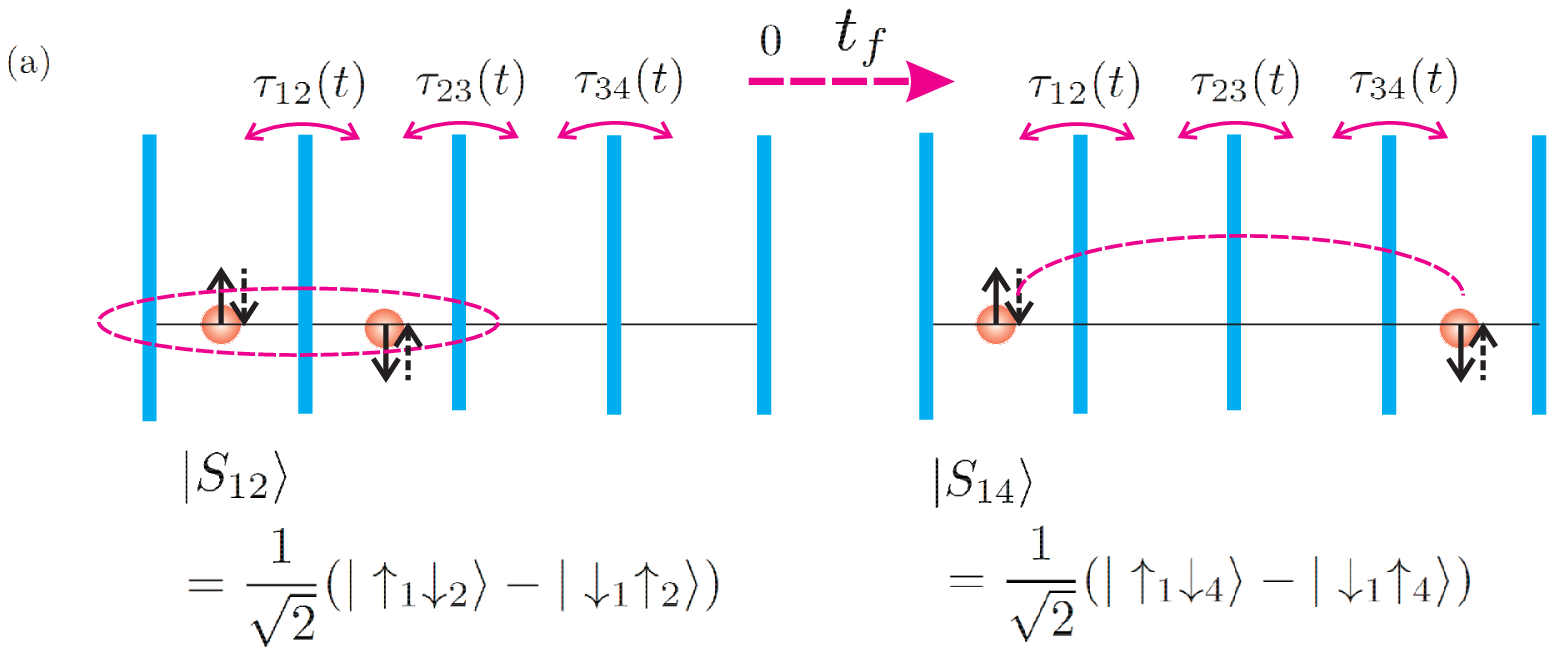}}
\scalebox{0.45}[0.45]{\includegraphics{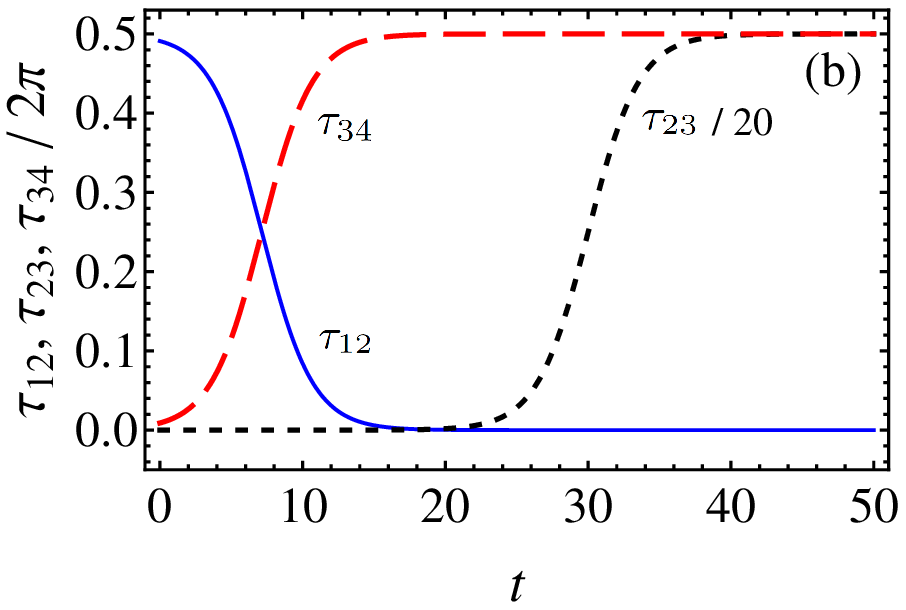}}
\scalebox{0.45}[0.45]{\includegraphics{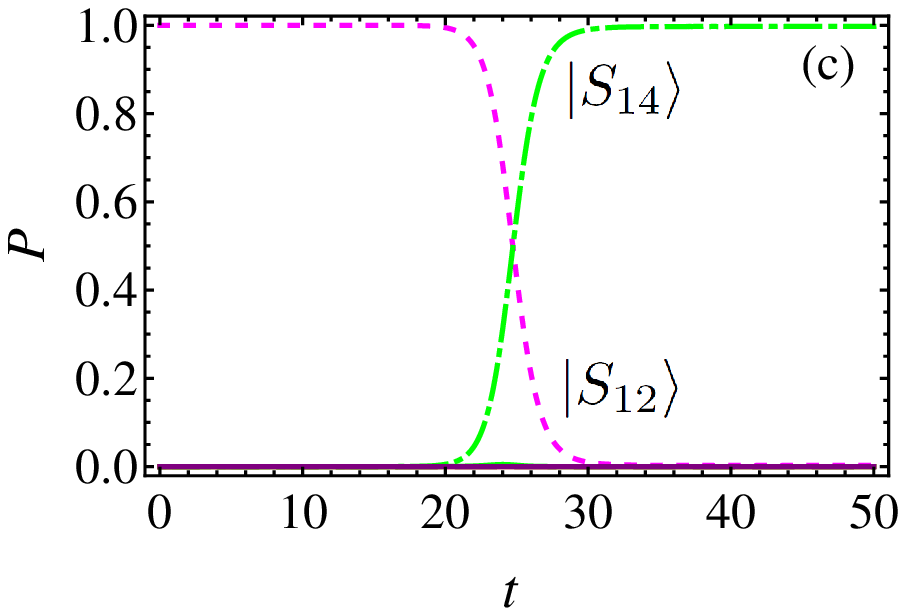}}
\caption{\label{4dots} (a)
Schematic diagram of CTAP from
$|S_{12}\rangle$ to $|S_{14}\rangle$ at $t_f= 50$ in a QQD by applying
(b) $\tau_{12}$, $\tau_{23}$ and $\tau_{34}$ [Eq.~(\ref{4dots-pulses-2})]. (c) $|S_{12}\rangle$ is transferred adiabatically to $|S_{14}\rangle$, where $U_0=1200\pi \sim 3$meV. }
\end{center}
\end{figure}

\textit{CTAP of entangled states in quadruple QDs.}---%
The case of a quadruple QD (QQD) deserves special attention, because the
traditional CTAP protocol for one electron requires an odd number of sites \cite{CTAP-QD}.
For two-electron entanglement transfer, by contrast, we find that this restriction
does not apply. Furthermore, we find out that there are two different DSs instead of one. Each of them allows for different transfer protocols. One of the two DS 
reads $|\phi^T_{DS,1}\rangle = \left[\tau_{12}, 0,
-\tau_{23}, 0,  0, \tau_{34}\right]^T/\nu_1$ for triplets, where $\tau_{12}$ and $\tau_{34}$ correspond to $\tau_{12}$ and $\tau_{23}$ for a TQD [Eq.~(\ref{3dots-pulses})] and $\tau_{23}$ is chosen as
\begin{equation}
\label{4dots-pulses}
\tau_{23} = \frac{3\tau_0}{2} \exp\left[-\frac{(t - t_f/2)^2}{\sigma^2}\right],
\end{equation}
where $\sigma = 3t_f/16$.  Then the population evolves via the DS
$|\phi^T_{DS,1}\rangle$ from $|T^0_{12}\rangle$ to $|T^0_{34}\rangle$, as shown in Fig. \ref{R-CTAP-4dots-1}. Similarly, a singlet is transferred provided that $U_0\gg |\tau^\text{max}_{ij}|^2$.
The other DS in a QQD is
\beqa
|\phi^T_{DS,2}\rangle = \frac{1}{\nu_2}\left[{\tau_{12}^2 - \tau_{34}^2, 0, -\tau_{12} \tau_{23}, \tau_{23} \tau_{34}, 0, 0}\right]^T,
\eeqa
where $\nu_2= \sqrt{(\tau_{12}^2 - \tau_{34}^2)^2 +\tau_{12}^2
\tau_{23}^2+\tau_{23}^2 \tau_{34}^2}$. Long-range entanglement transfer can
be adiabatically realized by coherently moving one electron along the DS
$|\phi^T_{DS,2}\rangle$. Specifically, $|S_{12}\rangle$ can be transferred to $|S_{14}\rangle$, as shown in Fig. \ref{4dots}(a), where the conditions $\tau_{23}(0) = \tau_{34}(0) = 0$,
$\tau_{12}(t_f) = 0$ and $\tau_{23}(t_f) \gg \tau_{34}(t_f)$ are satisfied.
To fulfill the above requirements, we choose $\tau_{12}$, $\tau_{34}$
in the same form of $\tau_{12}$ and $\tau_{23}$ from
Eq.~\eqref{3dots-pulses}, respectively, where $\tau_0= \pi/2$, $b=t_f/7$,
and
\beqa
\label{4dots-pulses-2}
\tau_{23} &=& a_0 \tau_0 \left[\tanh\left(\frac{t-b_2}{c}\right) + 1 \right],
\eeqa
where $b_2 = 3t_f/5$, $a_0=20$ (Fig.~\ref{4dots}(b)). High
fidelity transfer can be obtained by tuning the ratio $|\tau_{23}(t_f) /
\tau_{34}(t_f)|$.  Increasing $\tau_{34}(t_f)$ by adjusting $a_0$, can
further improve the fidelity. However, the pulse intensities should be
sufficiently small, as otherwise we witness leakage to higher states.
For instance, for the amplitude $\tau_{23}(t_f) = 20 \pi \sim 12
$GHz, which is experimentally feasible \cite{12dots-Petta}, $F=0.998$ is reached (see Fig.~\ref{4dots} (c)) with $U_0=1200\pi\sim3$meV. It approaches $1$ as $U_0$ increases.


\begin{figure}[tb]
\begin{center}
\scalebox{0.45}[0.45]{\includegraphics{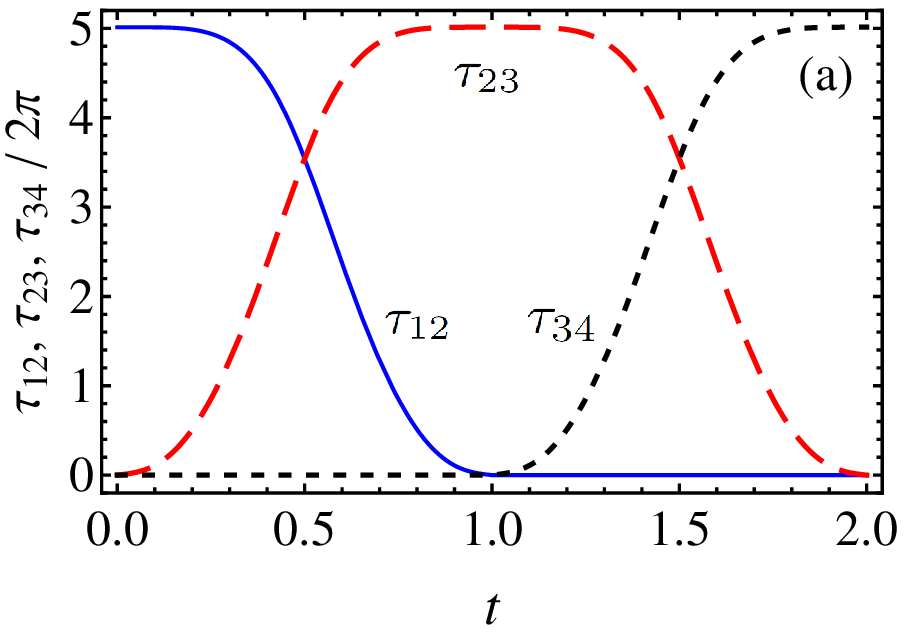}}
\scalebox{0.45}[0.45]{\includegraphics{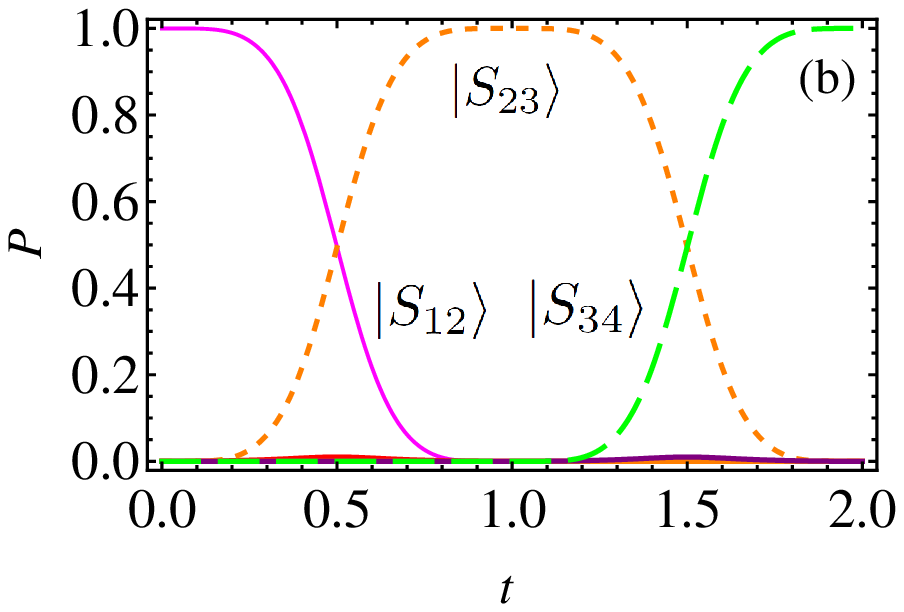}}
\caption{\label{STA-4dots}  STA transfer of a singlet in a
QQD with $t_f^{(4)}=2$, when $U_0 = 1400\pi \sim3.5$meV. (a)
With the pulses 
: $\tau^{(4)}_{12}$, $\tau^{(4)}_{23}$ and $\tau^{(4)}_{34}$, (b) the state is transferred from $|S_{12}\rangle$ to $|S_{34}\rangle$ via $|S_{23}\rangle$. Other states populations remain below $1\%$.
}
\end{center}
\end{figure}

\textit{Transfer of entangled states in longer arrays.}---%
We propose another protocol to transfer the entangled states in a
QD array with arbitrary length $n$, which is based on the protocol for a
TQD discussed above. $|T^0_{12}\rangle$ can be transferred to $|T^0_{(n-1)n}\rangle$
with the application of $n-2$ pulse sequences such that the total protocol
has the duration $(n-2)t_f$ and
\beqa
\tau^{(n)}_{k-2,k-1} &=& \tau^\text{CTAP}_{12}[t-(k-3)t_f],
\nonumber
\\
 \tau^{(n)}_{k-1,k} &=& \tau^\text{CTAP}_{23}[t-(k-3)t_f],
\label{pulses-ndots}
\eeqa
where $3\leq k \leq n$.  The pulses $\tau_{12}^\text{CTAP}$ and
$\tau_{23}^\text{CTAP}$ refer to the ones for TQD defined in Eq.~\eqref{3dots-pulses}.

\textit{Accelerating the transfer.}---%
In order to speed up the transfer, we use reverse engineering, a technique
of STA, which allows to design the pulses in order to reduce the transfer time.
Let us consider first TQD.
Since all onsite energies $\varepsilon_j=0$, we can
employ the ansatz $\Psi(t) = \cos\chi \cos\eta |1\rangle - i \sin\eta |2\rangle - \sin\chi \cos\eta |3\rangle$,
with $\chi(t)$ and $\eta(t)$ to be determined.
The conditions $\chi(0)=0$, $\chi(t_f)=\pi/2$, $\eta(0)=0$ and
$\eta(t_f)=0$ correspond to the initial and final state of the protocol,
while a smooth onset of the pulses is ensured by the conditions
$\dot\chi(0) = \dot\chi(t_f) = \ddot\chi(0) = \ddot\chi(t_f) =0$ and
$\dot\eta(0) = \dot\eta(t_f)=0$ \cite{three-level-I}.  To satisfy all the
above conditions, we introduce the scaled time $s=t/t_f$ and
choose $\chi(t) = \frac{\pi}{2} s  -\frac{15}{64}\sin(2\pi s) -
\frac{1}{192} \sin (6\pi s)$.  Besides the term linear in $t$, $\chi(t)$
consists of only the lowest odd Fourier components, which is a common
choice (``Gutman 1-3 trajectory'') for obtaining smooth pulses
\cite{Booktrajectory}.
In addition, the
function $\eta(t) = \arctan(\dot\chi / \alpha)$ is used, where the tunable
parameter $\alpha$ allows one to reduce the maximal occupation of the intermediate state to the value $P_{2}^\text{max} = \sin^2\eta(t_0) =
\dot\chi^2(t_0)/[\dot\chi^2(t_0)+\alpha^2]$ where $t_0 = t_f/2$.
Inserting the ansatz into the time-dependent
Schr\"{o}dinger equation, we obtain
\beqa
\label{pulses-STA}
\tau_{12}^\text{STA}(t) &=& \dot\eta \cos\chi + \dot\chi \cot\eta \sin\chi,
\nonumber
\\
\tau_{23}^\text{STA}(t) &=& -\dot\eta \sin\chi + \dot\chi \cot\eta \cos\chi,
\eeqa
for reversely engineering the pulses.

Figures~\ref{TS-3dots} compare the CTAP and STA protocols for a triplet transfer.  The operation time required is $t_f=1$, i.e.\
$50$ times shorter than the one obtained with CTAP, which corresponds to
$t_f=5.2$ns when $\tau^\text{max}_{ij} =5\mu$eV. The designed pulses in
Eq.~\eqref{pulses-STA} enable to speed-up the singlet transfer as well,
but requires that DOST are energetically distant from
the single occupied ones. 
For $U_0 = 1400\pi\sim 3.5$ meV, the transfer
from $|S_{12}\rangle$ to $|S_{23}\rangle$ occurs with $F>0.999$, while the
occupation of $|S_{13}\rangle$ stays below $1\%$ during the whole process.
A more detailed discussion concerning interaction can be found in \cite{supple}.

For the extension to an arbitrarily long QD array, we combine the ideas of CTAP in
longer arrays and the present STA pulses, i.e., we replace on the
right-hand side of Eq.~\eqref{pulses-ndots} the pulses in
Eq.~\eqref{pulses-STA}. In a QQD, the dynamics for the resulting STA protocol is
shown in Fig.~\ref{STA-4dots} for the singlet transfer. Our benchmark is
again the fidelity which should lie above the threshold $F>0.999$.  This
can be achieved with the pulses as in Eq. \ref{pulses-ndots}, where $\tau^{\textrm{CTAP}} _{ij}$ are substituted by $\tau^{\textrm{STA}}_{ij}$,
and the interaction $U_0=1200\pi\sim3$meV.

\textit{Robustness with respect to dephasing}.---%
So far, we have considered purely coherent quantum dynamics with perfect
pulses.  Experiments may operate under less ideal conditions.  We
therefore extend our numerical studies to the presence of decoherence
stemming from substrate phonons and fluctuations of the pulse strength.

Let us assume that each QD is coupled to a separate environment that
creates quantum noise coupled to the onsite energy.  Then our Hamiltonian must be extended by macroscopic
number of bosonic modes and a coupling Hamiltonian
$H_{\textrm{c}}=\sum_{j,\lambda}(n_{j,\uparrow}+n_{j,\downarrow})
(g_{j\lambda}a^{\dag}_{j\lambda} +g_{j\lambda}^* a_{j\lambda})$ where $a_{j\lambda}$ is the
annihilation operator of mode $\lambda$ coupled to QD $j$ with strength
$g_{j\lambda}$.  By standard techniques, one eliminates the bath within second
order perturbation theory and for an Ohmic spectral density of the bath
modes \cite{Breuer-book}.  After a Born-Markov and a rotating-wave approximation, one
finds the master equation
$\dot\rho = -i[H(t), \rho] -\sum_{j} \Gamma_j \left[n_{j\uparrow}+n_{j\downarrow},\left[n_{j\uparrow}+n_{j\downarrow}, \rho\right]\right]/2,$
with the dephasing rates $\Gamma_j$ which in our numerical studies assume
to be all equal.

\begin{figure}[tb]
\begin{center}
\scalebox{0.45}[0.45]{\includegraphics{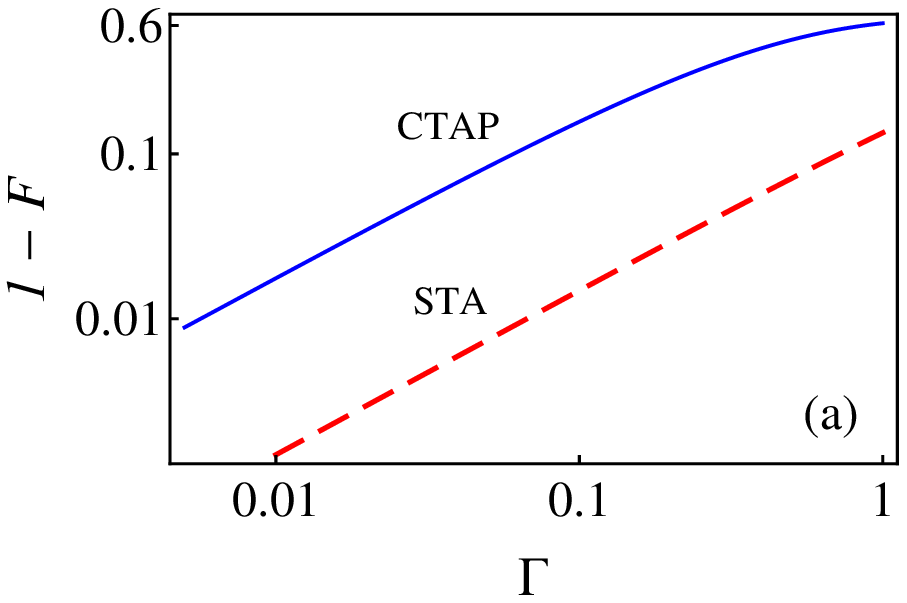}}
\scalebox{0.45}[0.45]{\includegraphics{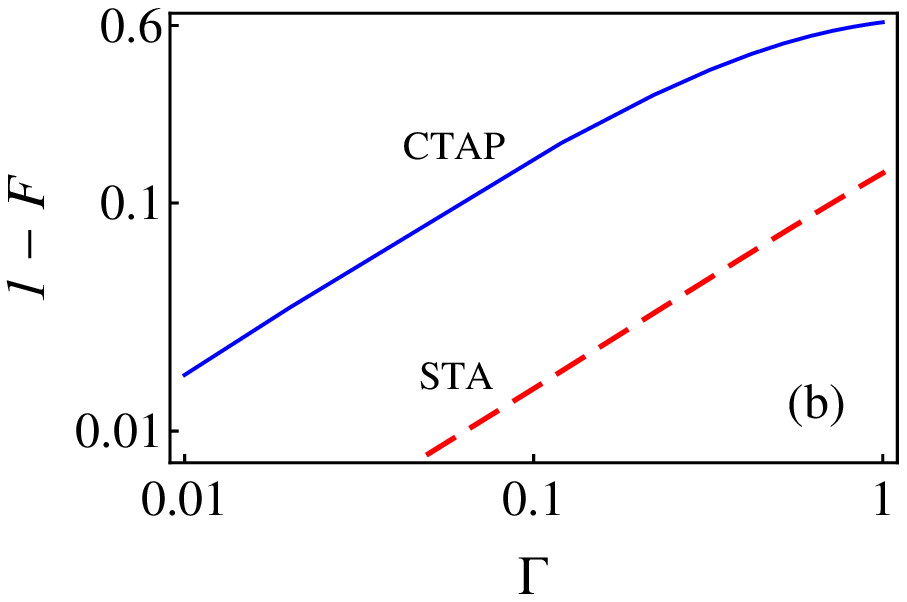}}
\\
\scalebox{0.45}[0.45]{\includegraphics{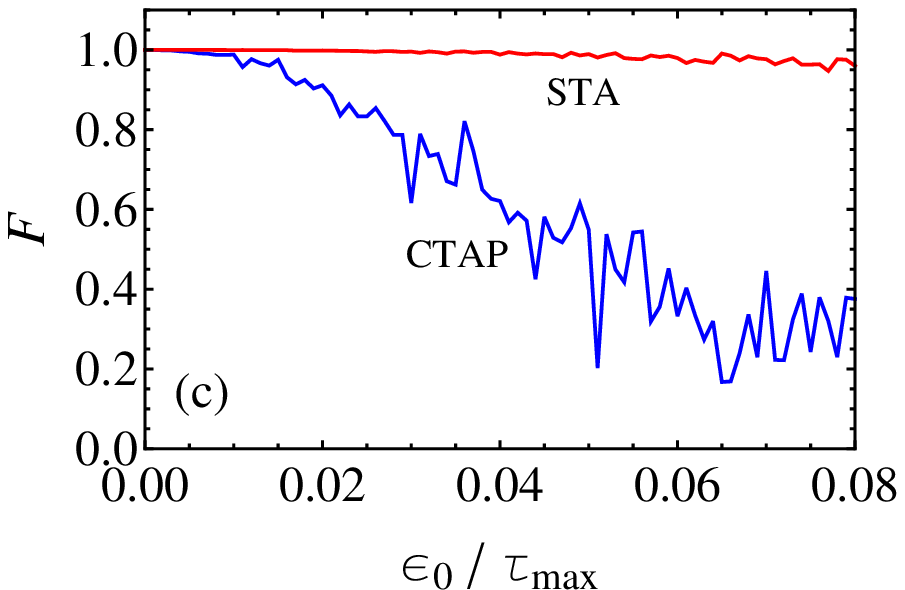}}
\scalebox{0.45}[0.45]{\includegraphics{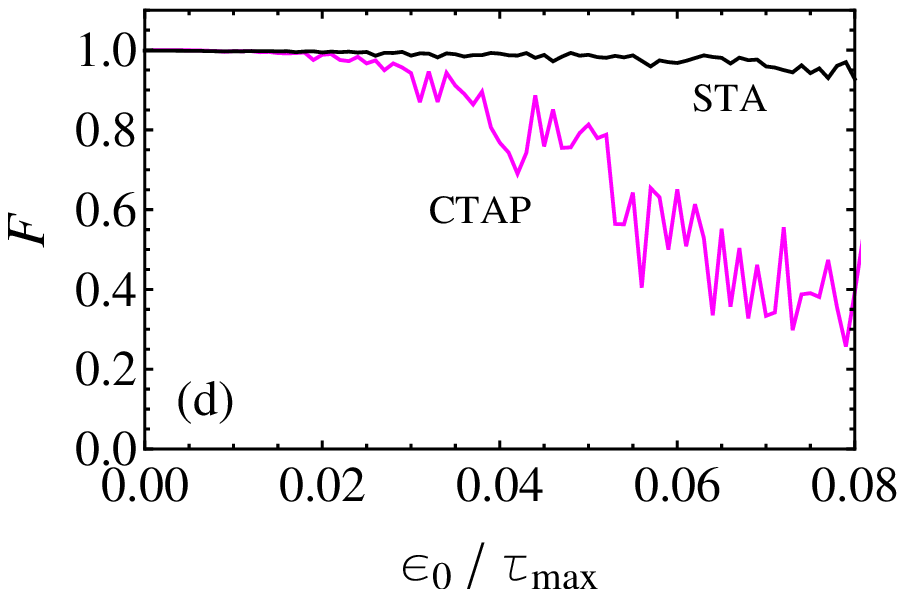}}
\caption{\label{3dots-noise-fluctuations} $1-F$, for the
triplet (a) and the singlet transfer (b) in a TQD versus the
dephasing rate $\Gamma$ in units of $1.92 \times 10^8 s^{-1}$, $U_0=1400\pi
\sim 3.5$ meV, for STA ($t_f=1$) and CTAP ($t_f=50$).
Fidelity of the triplet (c) and the singlet transfer (d) versus
random fluctuations in the electric pulses, by using STA ($t_f=1$) and CTAP
($t_f=50$), where $d = 35$, $\xi = t_f / 100$.}
\end{center}
\end{figure}

Figures~\ref{3dots-noise-fluctuations}(a,b) show the lack of fidelity $1-F$
of the protocol for the triplet and the singlet.
It turns out that it is proportional to the dephasing
rate, $1-F \propto \Gamma$.  For the singlet case, it is slightly larger,
which can be explained by the additional decay channel to DOST.  In both cases, the STA pulses perform significantly
better than the CTAP pulses.

Regarding pulse amplitudes fluctuations, we consider $\tilde{\tau}_{12}=\tau_{12} + \epsilon_{12}$,
$\tilde{\tau}_{23} = \tau_{23}+ \epsilon_{23}$, where $\epsilon_{12}$ and
$\epsilon_{23}$ are independent fluctuations.  We model them as Gaussian
pulses with strength $\epsilon_0$ at times $t_l$ and sign $\sigma_l=\pm1$,
such that their average vanishes.  Thus, $\epsilon = \epsilon_0
\sum_{l=1}^N s_l f(t-t_l)$ and $f(t-t_l) = \exp[-(t-t_l)^2/\xi^2]$.
The effects coming from the
noise are mainly related with the density of $t_l$ points, $d = N/t_f$, and
the width of distribution $\xi$, where $N$ are the total number of $t_l$ points. For the same value of fluctuation density
$d$, shorter $t_f$ results in less points of fluctuations. On the other
hand, a narrow width $\xi$ leads to more stable pulses. We compare the
dependence of $F$ on the amplitude  fluctuations by using CTAP
and STA protocols in a TQD, respectively. As shown in
Figs.~\ref{3dots-noise-fluctuations}(c,d), the transfer fidelities for
both triplet and singlet states are higher by using STA.
A further noise source may be the hyperfine interaction with the nuclear
spins, as discussed in \cite{supple}.

\textit{Conclusions.}---%
We have proposed a novel CTAP scheme for the long-range transfer of spin entangled states in QDs and derived for it a speed-up via STA.
The protocol works with rather large fidelity for both spin singlets and
triplets. Importantly, simultaneous transport of two particle spins from one edge to the other one of the atomic array is achieved while preserving their entanglement. While CTAP is designed for
slow operations, the operation of the STA is significantly faster, in our
numerical studies typically by a factor $50$.

When the coherent time evolution is affected by the
interaction with an environment, the STA protocol provides clear
advantages, because the reduced operation time make it less sensitive to
decoherence. Fluctuations of the pulse strength lead to similar conclusions.  For CTAP,
already fluctuations of the pulse strength below 1\% are noticeable, while
the STA scheme is fault tolerant towards much larger imperfections.

The precise control of electric pulses in the present experiments will
allow the implementation of the protocols presented here. In fact, the transfer
of quantum states between distant sites with high fidelity, could be
experimentally implemented not only in QDs but also in different physical
systems such as cold atoms or photonic crystals, which are of significance
for large-scale quantum information processing.

\begin{acknowledgments}
This work is supported by the Spanish Ministry of Science, Innovation, and
Universities via Grant No.\ MAT2017-86717-P, NSFC (11474193), SMSTC
(18010500400 and 18ZR1415500), and the Program for Eastern Scholar. Y.B.\
thanks the Juan de la Cierva Program of the Spanish MINECO.
\end{acknowledgments}

\end{document}


\title{Supplemental Material for \\
``Spin entangled state transfer in quantum dot arrays: Coherent adiabatic and speed-up protocols''}

\author{Yue Ban}
\affiliation{Instituto de Ciencia de Materiales de Madrid, CSIC, C/ Sor Juana In\'{e}s de la Cruz 3, E-28049 Madrid, Spain}
\affiliation{School of Materials Science and Engineering, Shanghai University, 200444 Shanghai, People's Republic of China}

\author{Xi Chen}
\affiliation{Department of Physics, Shanghai University, 200444 Shanghai, People's Republic of China}

\author{Sigmund Kohler}
\author{Gloria Platero}
\affiliation{Instituto de Ciencia de Materiales de Madrid, CSIC, C/ Sor Juana In\'{e}s de la Cruz 3, E-28049 Madrid, Spain}

\maketitle

\section{Two-electron entangled states transfer in a TQD}\label{sec:TQD}

In this section, we derive the Hamiltonian for two-electron entangled
states transfer in a TQD.  Our aim is to transfer the triplet state
$|T^0\rangle$ or the singlet state $|S\rangle$ from one edge to the other
one of a TQD. The triplet basis consists of three states
\beqa
\label{Triplet-3QD}
\nonumber
|1\rangle &=& |T^0_{12}\rangle =\frac{1}{\sqrt{2}}
(|{\uparrow}_1{\downarrow}_2\rangle +|{\downarrow}_1{\uparrow}_2\rangle),
\\
\nonumber
|2\rangle &=& |T^0_{13}\rangle =\frac{1}{\sqrt{2}}
(|{\uparrow}_1{\downarrow}_3\rangle +|{\downarrow}_1{\uparrow}_3\rangle),
\\
|3\rangle &=& |T^0_{23}\rangle =\frac{1}{\sqrt{2}}
(|{\uparrow}_2{\downarrow}_3\rangle +|{\downarrow}_2{\uparrow}_3\rangle),
\eeqa
where the subscripts label the dots occupied by electrons. In contrast, the
Hilbert space for the singlets consists of three single occupied and three
double occupied states:
\beqa
\label{Singlet-3QD}
\nonumber
|1\rangle &=& |S_{11}\rangle =|{\uparrow}_1{\downarrow}_1\rangle,
\\
\nonumber
|2\rangle &=& |S_{12}\rangle = \frac{1}{\sqrt{2}}
(|{\uparrow}_1{\downarrow}_2\rangle -|{\downarrow}_1{\uparrow}_2\rangle),
\\
\nonumber
|3\rangle &=& |S_{22}\rangle = |{\uparrow}_2\downarrow_2\rangle,
\\
\nonumber
|4\rangle &=& |S_{13}\rangle =\frac{1}{\sqrt{2}}
(|{\uparrow}_1{\downarrow}_3\rangle -|{\downarrow}_1{\uparrow}_3\rangle),
\\
\nonumber
|5\rangle &=& |S_{23}\rangle =\frac{1}{\sqrt{2}}
(|{\uparrow}_2{\downarrow}_3\rangle -|{\downarrow}_2{\uparrow}_3\rangle),
\\
|6\rangle &=& |S_{33}\rangle =|{\uparrow}_3{\downarrow}_3\rangle.
\eeqa
\begin{widetext}
The Hamiltonian expanded in the basis of the triplet states reads
\beqa
H_T = \left[\begin{array}{ccc}
\varepsilon_1 +\varepsilon_2  & \tau_{23} & 0
\\
\tau_{23}  & \varepsilon_1 + \varepsilon_3  & \tau_{12}
\\
0 & \tau_{12} & \varepsilon_2 +\varepsilon_3
\end{array}
\right] .
\label{HT}
\eeqa
Accordingly, the Hamiltonian in the basis of the singlet states becomes
\beqa
\label{HS}
H_S =
\nonumber
 \left[\begin{array}{cccccc}
 U_0+2\varepsilon_1 & \sqrt{2} \tau_{12} & 0 & 0 & 0 & 0
\\
\sqrt{2} \tau_{12}  &   \varepsilon_1 +\varepsilon_2 & \sqrt{2}\tau_{12} & \tau_{23} & 0 & 0
\\
0 & \sqrt{2}\tau_{12} & U_0 +2 \varepsilon_2 & 0 & \sqrt{2} \tau_{23} & 0
\\
0 & \tau_{23} & 0 & \varepsilon_{1} + \varepsilon_3  & \tau_{12} & 0
\\
0 & 0 & \sqrt{2}\tau_{23} & \tau_{12} & \varepsilon_2 + \varepsilon_3 & \sqrt{2}\tau_{23}
\\
0 & 0 & 0 & 0 & \sqrt{2}\tau_{23} & U_0+2\varepsilon_3
\end{array}\right].
\eeqa
By using the adiabatic elimination for $H_S$, i.e., we assume that the
first time derivative of the wavefunction of double occupied states is
zero, we can find the effective Hamiltonian of single occupied states
including implicitly the effects of the double occupied states
\beqa
\label{HS-effective}
H_S^{\textrm{eff}}= \left[\begin{array}{ccc}
-\frac{4 \tau^2_{12}}{U_0} & \tau_{23} & -\frac{2 \tau_{12}\tau_{23}}{U_0}
\\
\tau_{23}  & 0  & \tau_{12}
\\
-\frac{2 \tau_{12}\tau_{23}}{U_0} & \tau_{12} & -\frac{4 \tau^2_{23}}{U_0}
\end{array}
\right].
\eeqa
For $U_0 \gg |\tau^\text{max}_{ij}|^2$, this Hamiltonian becomes equal to
the one in Eq.~\eqref{HT} when setting all $\varepsilon_j=0$, such
that in this limit, the CTAP and STA pulses for the singlet states can also be
employed for the triplet states, as discussed in the main text.

With the same coupling pulses of $\tau_{12}$ and $\tau_{23}$ for the triplet transfer, the singlet state can be transferred  with high fidelity from $|S\rangle_{1,2}$ to $|S\rangle_{2,3}$, for $U_0 \gg |\tau^{max}_{ij}|^2$. We calculate the fidelity of the singlet transfer $F = |\langle S_{23}| \Psi(t_f) \rangle|^2$  for different values of the intradot Coulomb interaction $U_0$ for both CTAP Fig. \ref{S-F-U} (a) and STA \ref{S-F-U} (b). Both of them indicate that the fidelity of the singlet transfer reaches values $ F=0.999$ for  $U_0= 1400\pi$ $\sim 3.5$ meV which corresponds to Coulomb interaction values within the experimental ones for semiconductor QDs.

\begin{figure}[tb]
\begin{center}
\scalebox{0.55}[0.55]{\includegraphics{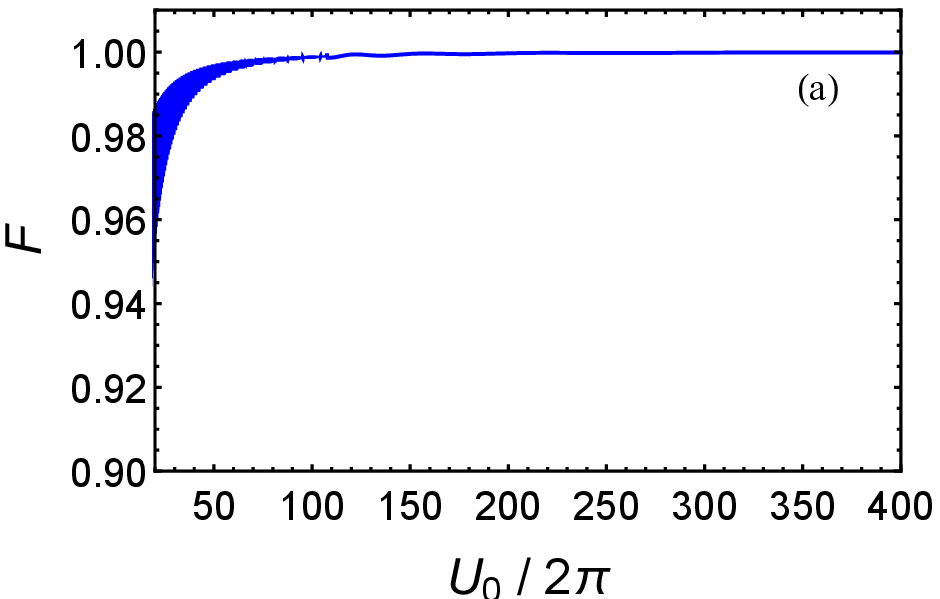}}
\scalebox{0.55}[0.55]{\includegraphics{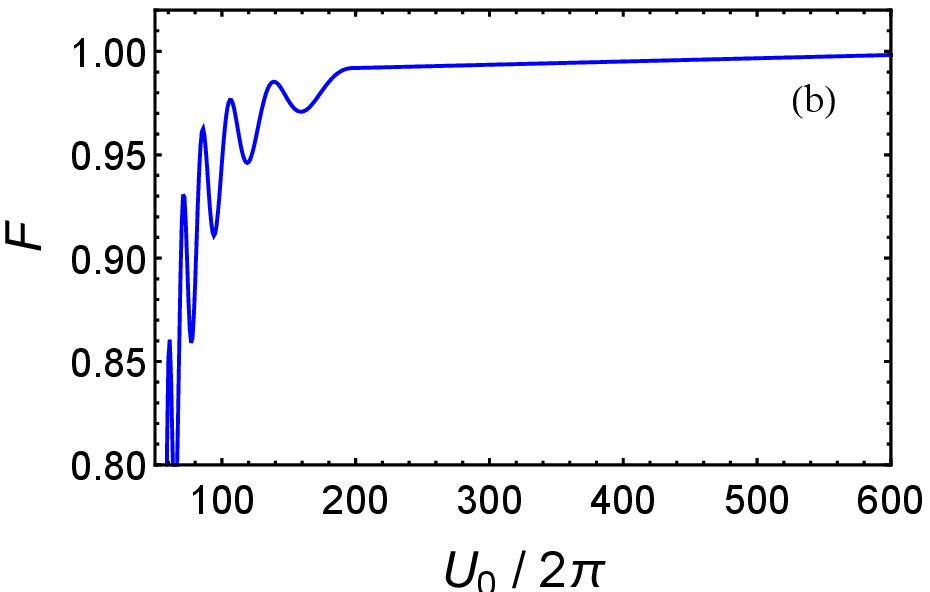}}
\caption{\label{S-F-U} Dependence of the fidelity $F$ for the singlet transfer in a TQD on $U_0$ by CTAP, where $t_f =50$ (a) and STA, where $t_f=1$ (b).}
\end{center}
\end{figure}

\section{Hamiltonian of two-electron triplet/singlet state in a quadruple QD}
\label{sec:4QD}
In this section, we derive the Hamiltonian of two-electron triplet/singlet
states in a quadruple QD.  After expanding the Hamiltonian in the basis of
the triplet states in a quadruple QD, where $|1\rangle =
|T^0_{12}\rangle$, $|2\rangle = |T^0_{13}\rangle$, $|3\rangle =
|T^0_{23}\rangle$, $|4\rangle = |T^0_{14}\rangle$, $|5\rangle =
|T^0_{24}\rangle$, $|6\rangle = |T^0_{34}\rangle$, we obtain the
Hamiltonian
\beqa
H_T =\left[\begin{array}{cccccc}
 \varepsilon_1 +\varepsilon_2 & \tau_{23} & 0 & 0 & 0 & 0
\\
\tau_{23}  & \varepsilon_1 + \varepsilon_3 & \tau_{12} & \tau_{34} & 0 & 0
\\
0 & \tau_{12} & \varepsilon_2 +\varepsilon_3 & 0 & \tau_{34} & 0
\\
0 & \tau_{34} & 0 & \varepsilon_{1} + \varepsilon_{4} & \tau_{12} & 0
\\
0 & 0 & \tau_{34} & \tau_{12} & \varepsilon_3 +\varepsilon_4 & \tau_{23}
\\
0 & 0 & 0 & 0 & \tau_{23} & \varepsilon_3 + \varepsilon_4
\end{array}\right].
\label{HT-4dots}
\eeqa
Setting all $\varepsilon_j=0$, we obtain after some algebra the two dark
states
\begin{align}
|\phi^T\rangle_{DS,1}
={}& \frac{1}{\nu_1}\left[\tau_{12}, 0,  -\tau_{23}, 0,  0, \tau_{34}\right]^T,
\\
|\phi^T\rangle_{DS,2}
={}& \frac{1}{\nu_2}\left[\tau_{12}^2 - \tau_{34}^2, 0, -\tau_{12}
\tau_{23}, \tau_{23} \tau_{34}, 0, 0\right]^T ,
\end{align}
where the normalizations $\nu_1= \sqrt{\tau_{12}^2 +\tau_{23}^2
+\tau_{34}^2}$ and with $\nu_2 = \sqrt{\tau_{23}^2 \tau_{34}^2 +
\tau_{12}^2 \tau_{23}^2 + (\tau_{12}^2 - \tau_{34}^2)^2}$.
By using these two dark states and appropriate pulses, one
can transfer adiabatically the triplet state from $|T^0_{12}\rangle$ to
$|T^0_{34}\rangle$ and to $|T^0_{14}\rangle$, respectively.

On the other hand, the singlet subspace is comprised by six single occupied
states and four double occupied ones, $|1\rangle = |S_{11}\rangle$,
$|2\rangle = |S_{12}\rangle$, $|3\rangle = |S_{22}\rangle$, $|4\rangle =
|S_{13}\rangle$, $|5\rangle = |S_{23}\rangle$, $|6\rangle =
|S_{33}\rangle$, $|7\rangle = |S_{14}\rangle$, $|8\rangle =
|S_{24}\rangle$, $|9\rangle = |S_{34}\rangle$, $|10\rangle =
|S_{44}\rangle$, corresponding to the Hamiltonian
\begin{eqnarray}
\label{HS-4dots}
H_S=\begin{pmatrix}
U_{0}+2\varepsilon_1 & \sqrt{2} \tau_{12} & 0 & 0 & 0 & 0 & 0 & 0 & 0 & 0
\\
\sqrt{2} \tau_{12}  &  \varepsilon_1 + \varepsilon_2 & \sqrt{2} \tau_{12} & \tau_{23} & 0 & 0 & 0 & 0 & 0 & 0
\\
0 & \sqrt{2}\tau_{12} & U_0+2\varepsilon_2 & 0 & \sqrt{2}\tau_{23} & 0 & 0 & 0 & 0 & 0
\\
0 & \tau_{23} & 0 & \varepsilon_{1} + \varepsilon_3 & \tau_{12} & 0 & \tau_{34} & 0 & 0 & 0
\\
0 & 0 & \sqrt{2} \tau_{23} & \tau_{12} & \varepsilon_2 + \varepsilon_3 & \sqrt{2}\tau_{23} & 0 & \tau_{34} & 0 & 0
\\
0 & 0 & 0 & 0 & \sqrt{2} \tau_{23} & U_0 + 2\varepsilon_3 & 0 & 0 & \sqrt{2}\tau_{34} & 0
\\
0 & 0 & 0 & \tau_{34} & 0 & 0 & \varepsilon_1 +\varepsilon_4 & \tau_{12} & 0 & 0
\\
0 & 0 & 0 & 0 & \tau_{34} & 0 & \tau_{12} & \varepsilon_2 + \varepsilon_4 & \tau_{23} & 0
\\
0 & 0 & 0 & 0 & 0 & \sqrt{2}\tau_{34} & 0 & \tau_{23} & \varepsilon_3 + \varepsilon_4 & \sqrt{2}\tau_{34}
\\
0 & 0 & 0 & 0 & 0 & 0 & 0 & 0 & \sqrt{2}\tau_{34} & U_0 +2\varepsilon_4
\end{pmatrix}.
\end{eqnarray}
\begin{figure}[tb]
\begin{center}
\scalebox{0.55}[0.55]{\includegraphics{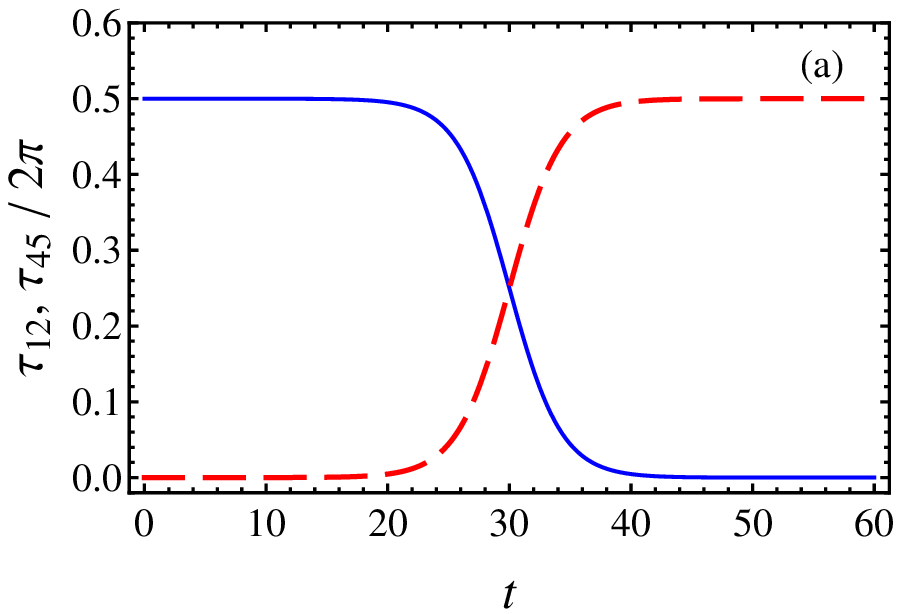}}
\scalebox{0.55}[0.55]{\includegraphics{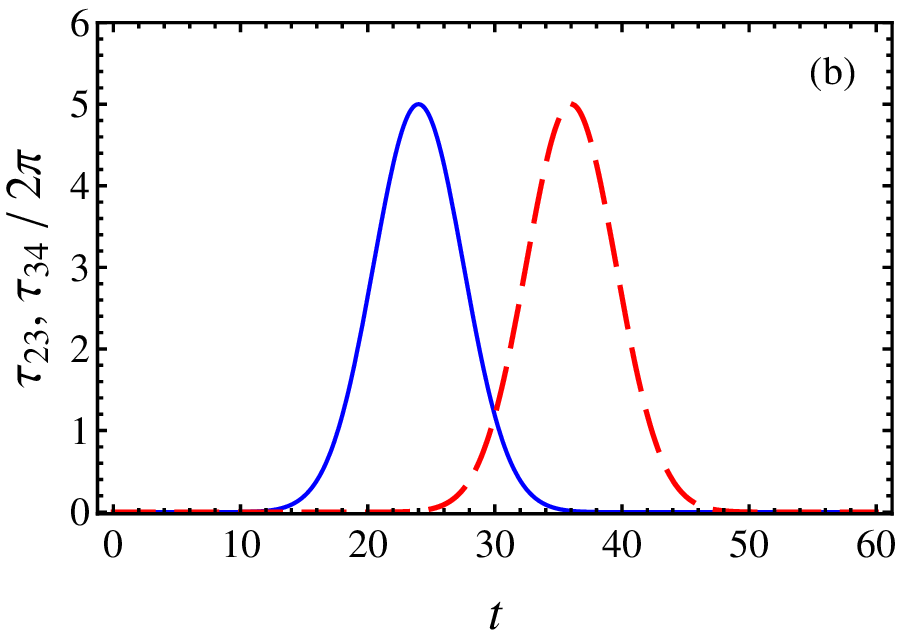}}
\\
\scalebox{0.55}[0.55]{\includegraphics{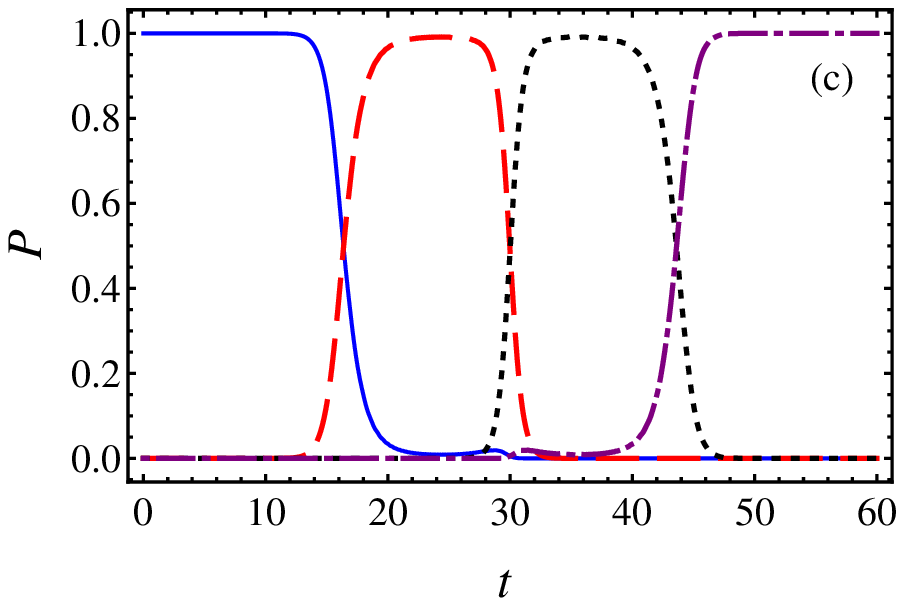}}
\scalebox{0.55}[0.55]{\includegraphics{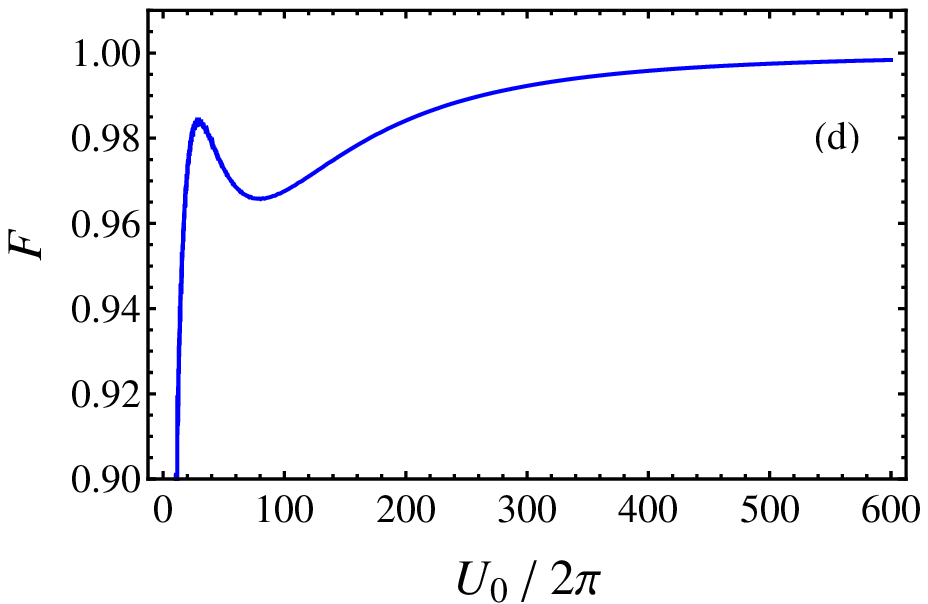}}
\caption{\label{R-CTAP-T-5dots} CTAP transfer of triplet/singlet in a
quintuple QD. With the applications of (a) $\tau_{12}$ (blue, solid),
$\tau_{45}$ (red, dashed) and (b) $\tau_{23}$ (blue, solid), $\tau_{34}$
(red, dashed), see Eq. (\ref{5dots-pulses}), (c) the triplet state is
transferred adiabatically from $|T^0_{12}\rangle$ (blue, solid) to
$|T^0_{45}\rangle$ (purple, dot-dashed) at $t_f = 60 \sim312$ ns, with
population excitation of $|T^0_{23}\rangle$ (red, dashed) and
$|T^0_{34}\rangle$ (black, dotted) at intermediate times. (d) Dependence
of the fidelity of a singlet transfer on $U_0$.}
\end{center}
\end{figure}

In order to see the effects of double occupied states during the transfer
in a quadruple QD, we also do the adiabatic elimination for $H_S$, and find
the Hamiltonian with zero detuning
\begin{eqnarray}
\label{HS-4dots-eff}
H_S=\begin{pmatrix}
-\frac{4\tau_{12}^2}{U_0} & \tau_{23} & -\frac{2\tau_{12}\tau_{23}}{U_0} & 0 & 0 & 0
\\
\tau_{23} & 0 & \tau_{12} & \tau_{34} & 0 & 0
\\
-\frac{2\tau_{12}\tau_{23}}{U_0} & \tau_{12} & -\frac{4\tau_{23}^2}{U_0} & 0 & \tau_{34} & -\frac{2\tau_{23}\tau_{34}}{U_0}
\\
0 & \tau_{34} & 0 & 0 & \tau_{12} & 0
\\
0 & 0 & \tau_{34} & \tau_{12} & 0 & \tau_{23}
\\
0 & 0 & -\frac{2\tau_{23}\tau_{34}}{U_0} & 0 & \tau_{23} & -\frac{4\tau_{34}^2}{U_0}
\end{pmatrix}.
\end{eqnarray}
As for the TQD, we take the limit $U_0\gg |\tau_{ij}^\text{max}|^2$, such
that $H_S$ becomes identical to $H_T$ in Eq.~\eqref{HT-4dots} with all
$\varepsilon_j=0$.

\section{CTAP transfer of the two-electron triplet/singlet state in a quintuple QD}
\label{sec:5QD}
In a quintuple QD, the number of states of the triplet and the singlet basis increases significantly.
The ten-state triplet Hamiltonian becomes
\beqa
\label{HT-5dots}
H_T = \left[\begin{array}{cccccccccc}
\varepsilon_1 + \varepsilon_2 & \tau_{23} & 0 & 0 & 0 & 0 & 0 & 0 & 0 & 0
\\
\tau_{23} & \varepsilon_{1} + \varepsilon_3 & \tau_{12} & \tau_{34} & 0 & 0 & 0 & 0 & 0 & 0
\\
0 & \tau_{12} & \varepsilon_2 + \varepsilon_3 & 0 & \tau_{34} & 0 & 0 & 0 & 0 & 0
\\
0 & \tau_{34} & 0 & \varepsilon_1 +\varepsilon_4 & \tau_{12} & 0 & \tau_{45} & 0 & 0 & 0
\\
0 & 0 & \tau_{34} & \tau_{12} & \varepsilon_2 + \varepsilon_4 & \tau_{23} & 0 & \tau_{45} & 0 & 0
\\
0 & 0 & 0 & 0 & \tau_{23} & \varepsilon_3 + \varepsilon_4 & 0 & 0 & \tau_{45} & 0
\\
0 & 0 & 0 & \tau_{45} & 0 & 0 & \varepsilon_1 + \varepsilon_5 & \tau_{12} & 0 & 0
\\
0 & 0 & 0 & 0 & \tau_{45} & 0 & \tau_{12} & \varepsilon_2 + \varepsilon_5 & \tau_{23} & 0
\\
0 & 0 & 0 & 0 & 0 & \tau_{45} & 0 & \tau_{23} & \varepsilon_3 + \varepsilon_5 & \tau_{34}
\\
0 & 0 & 0 & 0 & 0 & 0 & 0 & 0 & \tau_{34} & \varepsilon_4 + \varepsilon_5
\end{array}
\right],
\eeqa
and the fifteen-state singlet Hamiltonian is
\beqa
\label{HS-5dots}
H_S = \left[\begin{smallmatrix}
U_{0}+2\varepsilon_1 & \sqrt{2} \tau_{12} & 0 & 0 & 0 & 0 & 0 & 0 & 0 & 0 & 0 & 0 & 0 & 0 & 0
\\
\sqrt{2} \tau_{12}  &  \varepsilon_1 + \varepsilon_2 & \sqrt{2} \tau_{12} & \tau_{23} & 0 & 0 & 0 & 0 & 0 & 0 & 0 & 0 & 0 & 0 & 0
\\
0 & \sqrt{2}\tau_{12} & U_0+2\varepsilon_2 & 0 & \sqrt{2}\tau_{23} & 0 & 0 & 0 & 0 & 0 & 0 & 0 & 0 & 0 & 0
\\
0 & \tau_{23} & 0 & \varepsilon_{1} + \varepsilon_3 & \tau_{12} & 0 & \tau_{34} & 0 & 0 & 0 & 0 & 0 & 0 & 0 & 0
\\
0 & 0 & \sqrt{2} \tau_{23} & \tau_{12} & \varepsilon_2 + \varepsilon_3 & \sqrt{2}\tau_{23} & 0 & \tau_{34} & 0 & 0 & 0 & 0 & 0 & 0 & 0
\\
0 & 0 & 0 & 0 & \sqrt{2} \tau_{23} & U_0 + 2\varepsilon_3 & 0 & 0 & \sqrt{2}\tau_{34} & 0 & 0 & 0 & 0 & 0 & 0
\\
0 & 0 & 0 & \tau_{34} & 0 & 0 & \varepsilon_1+\varepsilon_4 & \tau_{12} & 0 & 0 & \tau_{45} & 0 & 0 & 0 & 0
\\
0 & 0 & 0 & 0 & \tau_{34} & 0 & \tau_{12} & \varepsilon_2 + \varepsilon_4 & -\tau_{23} & 0 & 0 & \tau_{45} & 0 & 0 & 0
\\
0 & 0 & 0 & 0 & 0 & \sqrt{2}\tau_{34} & 0 & \tau_{23} & \varepsilon_3 + \varepsilon_4 & \sqrt{2}\tau_{34} & 0 & 0 & \tau_{45} & 0 & 0
\\
0 & 0 & 0 & 0 & 0 & 0 & 0 & 0 & \sqrt{2}\tau_{34} & U_0 +2\varepsilon_4 & 0 & 0 & 0 & \sqrt{2}\tau_{45} & 0
\\
0 & 0 & 0 & 0 & 0 & 0 & \tau_{45} & 0 & 0 & 0 & \varepsilon_1 + \varepsilon_5 & \tau_{12} & 0 & 0 & 0
\\
0 & 0 & 0 & 0 & 0 & 0 & 0 & \tau_{45} & 0 & 0 & \tau_{12} & \varepsilon_2 + \varepsilon_5 & \tau_{23} & 0 & 0
\\
0 & 0 & 0 & 0 & 0 & 0 & 0 & 0 & \tau_{45} & 0 & 0 & \tau_{23} & \varepsilon_3 + \varepsilon_5 & \tau_{34} & 0
\\
0 & 0 & 0 & 0 & 0 & 0 & 0 & 0 & 0 & \sqrt{2}\tau_{45} & 0 & 0 & \tau_{34} & \varepsilon_4 + \varepsilon_5 & \sqrt{45}\tau_{45}
\\
0 & 0 & 0 & 0 & 0 & 0 & 0 & 0 & 0 & 0 & 0 & 0 & 0 & \sqrt{2} \tau_{45} & 2\varepsilon_5+ U_0
\end{smallmatrix}\right].
\eeqa

Similar to the case of TQD, we can find the dark state of $H_T$ with zero
detuning,
\beqa
\label{5dots-dark-state}
\frac{1}{\nu_1}[\tau_{12}, 0, -\tau_{23}, 0, 0, \tau_{34}, 0, 0, 0, -\tau_{45}]^T,
\eeqa
where $\nu_1=\sqrt{\tau_{12}^2+\tau_{23}^2+\tau_{34}^2 + \tau_{45}^2}$, such
that adiabatic triplet transfer from $|T^0_{12}\rangle =
[1,0,0,0,0,0,0,0,0,0]^T$ to $|T^0_{45}\rangle = [0,0,0,0,0,0,0,0,0,1]^T$
can be achieved by applying the pulses
\beqa
\label{5dots-pulses}
\tau_{12} &=& -\frac{\tau_0}{2} \left[\tanh\left(\frac{t-b}{c}\right) - 1 \right],
\\
\tau_{45} &=& \frac{\tau_0}{2}\left[\tanh\left(\frac{t-b}{c}\right) + 1 \right],
\nonumber
\\
\tau_{23} &=& 10 \tau_0 \exp\left[-\frac{(t-t_f/2 +\tau)^2}{\sigma^2}\right],
\nonumber
\\
\tau_{34} &=& 10 \tau_0 \exp\left[-\frac{(t-t_f/2 -\tau)^2}{\sigma^2}\right],
\nonumber
\eeqa
where the parameters are chosen as $b= t_f/2$, $c = t_f /14$, $\tau_0 =
\pi$, $\tau = t_f/10$, $\sigma = t_f/12$ and $t_f =60 \sim 312$ns, in order
to obtain the transfer fidelity $F>0.999$ at $t=t_f$.
Figs.~\ref{R-CTAP-T-5dots}(a) and (b) show the designed pulses. The triplet state
$|T^0_{12}\rangle$ is transferred into $|T^0_{45}\rangle$,
with populating $|T^0_{23}\rangle$ and $|T^0_{34}\rangle$ intermediately,
indicated in Fig.~\ref{R-CTAP-T-5dots}(c). Fig.~\ref{R-CTAP-T-5dots}(d)
shows the dependence of the singlet state transfer fidelity with the
Hubbard on-site interaction. Here, we choose $\tau_{23}^{\textrm{max}} =
\tau_{34}^{\textrm{max}} \sim 6$ GHz and $t_f =60\sim 312$ ns, which
gives rise to $F=0.999$ for the triplet transfer. Using these pulses, by setting $U_0 > 1000 \pi
\sim 2.5$ eV, we can achieve the fidelity of the singlet transfer higher
than $0.999$.

\end{widetext}

\begin{table}[t]
\caption{\label{Transfer Fidelity}
Transfer fidelity of the triplet/singlet state in a TQD in the presence of
hyperfine interaction, with $T_1=50\mu$s and $T_2=1\mu$s.  The pulses
designed from CTAP, $t_f=50\sim 260$ns and STA, $t_f =1\sim 5.2$\,ns are
shown in Figs. 2(a) and (c) of the main text, respectively.}
\begin{ruledtabular}
\begin{tabular}{lccr}
 & Pulse duration & $F$ Triplet & $F$ Singlet \\
\hline
STA & $t_f =1$ & 0.995 & 0.988
 \\
\hline
CTAP & $t_f=50$ & 0.795 & 0.787
\end{tabular}
\end{ruledtabular}
\end{table}

\section{Singlet/triplet transfer in a TQD in the presence of a nuclear field.}
One of the main spin decoherence mechanisms in GaAs quantum dots is
hyperfine interaction. Here, we include in our analysis the effect of
hyperfine interaction between electron and nuclear spins. The effective
magnetic field $\bm{B}^N_j$ resulting from a random configuration of many
nuclear spins localized in each quantum dot,
affects the spin of the electrons, and also causes transitions between
singlet and triplet states. Now, we consider a Hamiltonian with both
singlets and triplets where we add also the parallel spin states
$|T^{+}_{ij}\rangle = |T^{{\uparrow}{\uparrow}}_{ij}\rangle$ and
$|T^{-}_{ij}\rangle = |T^{{\downarrow}{\downarrow}}_{ij}\rangle$.

Here, we consider zero detuning and an external magnetic field of several
hundreds of mT, i.e., much larger than the Overhauser field $B^N$  which is
of the order of $5$mT. We consider a phenomenological model in GaAs QDs
where we include in the time evolution of the density matrix equation a
phenomenological spin-flip rate.  As the difference between the
$|S\rangle$ and $|T^0\rangle$ energy levels is much smaller than the one
between $|S\rangle$ and $|T^{\pm}\rangle$, and the one between
$|T^0\rangle$ and $|T^{\pm}\rangle$, transitions occur mainly between
$|S_{ij}\rangle$ and $|T^0_{ij}\rangle$. Thus, we consider the Hilbert space which includes both subspaces and we consider a spin relaxation time of $T_1=50\mu$s and a spin decoherence time of $T_2=1\mu$s, (which are typical experimental values for GaAs QDs).
Then, we solve the reduced density matrix by using the pulses designed by CTAP and STA. We obtain the transfer fidelity for the singlet and triplet states, as shown in Table~\ref{Transfer Fidelity}.

Using the strategy of STA, we obtain high fidelity, even in the presence of
hyperfine interaction, for the transfer of both triplet and singlet states.
The reason is that decoherence is avoided to large extent by shortening the
operation time from $t_f=50$ ($260$ns) into $t_f=1$ ($5.2$ns), which is
much smaller than the spin decoherence time considered in our calculation
$T_2=1\mu$s.